# Identifying Malicious Web Domains Using Machine Learning Techniques with Online Credibility and Performance Data

Zhongyi Hu[1], Raymond Chiong[2], Ilung Pranata[2], Willy Susilo[3], and Yukun Bao[4]

[1]School of Information Management
Wuhan University, Wuhan, 430072, China
E-mail: Zhongyi.Hu@whu.edu.cn

[2]School of Design, Communication & Information Technology
The University of Newcastle, Callaghan, NSW 2308, Australia
E-mails: {Raymond.Chiong, Ilung.Pranata}@newcastle.edu.au

[3]School of Computing & Information Technology
The University of Wollongong, Wollongong, NSW 2252, Australia
E-mail: wsusilo@uow.edu.au

[4]School of Management
Huazhong University of Science and Technology, Wuhan, 430074, China
E-mail: yukunbao@hust.edu.cn

***Abstract*— Malicious web domains represent a big threat to web users' privacy and security. With so much freely available data on the Internet about web domains' popularity and performance, this study investigated the performance of well-known machine learning techniques used in conjunction with this type of online data to identify malicious web domains. Two datasets consisting of malware and phishing domains were collected to build and evaluate the machine learning classifiers. Five single classifiers and four ensemble classifiers were applied to distinguish malicious domains from benign ones. In addition, a binary particle swarm optimisation (BPSO) based feature selection method was used to improve the performance of single classifiers. Experimental results show that, based on the web domains' popularity and performance data features, the examined machine learning techniques can accurately identify malicious domains in different ways. Furthermore, the BPSO-based feature selection procedure is shown to be an effective way to improve the performance of classifiers.**

***Keywords—Malicious web domains; web security; classification; machine learning; feature selection***

## I. INTRODUCTION

The increased popularity and widespread use of Internet has made e-marketing an efficient way to promote products and services online. These days, organisations often share information about their products or services via email and social media. With an insurmountable amount of information from e-marketing, attacks through malicious URLs that trick an email recipient to an illegitimate site have become a common threat to web users' privacy and security. Two main types of malicious attacks are *phishing attacks* that deceive users into sharing passwords and their private information, as well as *malware attacks* that secretly access and infect users' computers by distributing viruses and malicious software. Malicious attacks can cause not just financial loss and privacy violation but also shatter the trust of web users in conducting e-commerce and online social activities, thereby indirectly affecting the effectiveness of e-marketing. To protect users from malicious attacks, a typical way is to prevent users from visiting malicious online links by providing timely warning against accessing content on malicious web domains. Being able to accurately identify malicious web domains has therefore become a matter of paramount importance.

Over the years, many anti-phishing toolbars [1, 2] have been developed, primarily by maintaining a user-verified blacklist. This technique detects phishing web domains by comparing the web domain a user visits with phishing web domains collected in the blacklist. However, due to the rapid increase of phishing web domains over time, it is not only challenging but almost impossible to maintain a perfect blacklist. Some studies have investigated and confirmed the ineffectiveness of various blacklist-based security toolbars in identifying phishing web domains (e.g., see [3]).

A more effective way to tackle the issue is through the use of machine learning. Many popular machine learning techniques have been successfully applied to identify phishing websites based on data features extracted from the sites. For example, Blum et al. [4] developed confidence-weighted classification for phishing URL detection based on lexical features. Ma et al. [5, 6] created classification algorithms to detect malicious URLs by looking at the lexical and host-based properties of suspicious URLs. Garera et al. [7] identified features from URLs to model a logistic regression classifier in order to distinguish a phishing URL from benign ones. In addition to lexical and host-based features, some studies also extracted features from web page content and used the features to train machine learning algorithms for malicious site identification [8-11]. For instance, Ludl et al. [12] investigated the feasibility of using properties of a web page, such as images, texts and links, to identify phishing pages based on decision trees. While these techniques have shown to have good accuracies in detecting phishing sites, they require extraction of features and comparison between the benign and malicious sites, which can be a tedious process.

For the drive-by-download type of (malware) attacks, a common way is to detect signatures of malicious sites by reviewing their code either manually or automatically. Although doing so can be simple and straightforward, it is sometimes ineffective as the code review process could be easily evaded by obfuscated code. Thus, researchers have turned to machine learning techniques for overcoming the problem by applying features extracted from the content or code of malicious web pages to build identification models. Canali et al. [10], for instance, proposed a filter called *Prophiler* to extract features from content of a page as well as from URLs, and from there they derived detection models with machine learning techniques for malicious site identification. Curtsinger et al. [13] proposed a Bayesian-based model using features of Javascript to identify syntax elements of malware sites. Similar to the identification of phishing web domains with content-based features, these approaches could be cumbersome as features need to be extracted and analysed from web content.

Given that different kinds of metrics for web domains are freely available and can be used to evaluate their credibility, this paper investigates the performance of machine learning techniques based on this type of widely available Internet data (i.e., the popularity and performance of web domains) to identify malicious web domains. In total, we consider 16 data features (see Section IV.A for details) and nine machine learning models, which include both single classifiers and ensemble models (see Sections III.A and III.B). To reduce the computational burden and further improve the performance of identification, a binary particle swarm optimisation (BPSO) based feature selection technique is implemented for the single classifiers to select a subset of the features rather than applying all of them (see Section III.C).

The rest of the paper is structured as follows. Section II reviews related studies in the area. In Section III, we describe the machine learning models and feature selection technique used. Experimental settings and results with discussions are presented in Section IV. Finally, Section V concludes this study with highlights of future research directions.

## II. Related Work

A number of studies in the literature have employed machine learning techniques to identify phishing websites. Ludl et al. [12] performed a study on the effectiveness of anti-phishing solutions using blacklists. They also investigated the feasibility of using properties of a web page, such as images, texts and links, to identify phishing pages. Two major findings from this study are 1) blacklist-based solutions are actually quite effective in protecting users from phishing, and 2) web page properties can be used to identify phishing. Ludl et al. used a total of 18 data features from web pages in their classification analysis. Xiang et al. [8, 9] proposed Cantina+, a feature-rich machine learning framework that adopts the HTML Document Object Model and search engine features to detect phishing sites. Cantina+ first applies two filters, i.e., hash-based and login form filters, to filter near-duplicate phishing websites. If a website is not identified as a phishing site by these filters, Cantina+ then tests the website using machine learning techniques with a total of 15 features. These features include URL, HTML and search engine based features. An additional feature, a domain top-page similarity feature, was later added to Cantina+ by Sanglerdsinlapachai and Rungsawang [14]. Whittaker et al. [15] developed a scalable machine learning classifier to maintain Google's phishing blacklist. Their system classifies web pages that are submitted by end users and/or collected from Gmail's spam filters. Similar to Cantina+, the proposed system adopts URL and HTML features. To reduce the number of features used by their machine learning software, the system only uses features of terms with the highest Term Frequency-Inverse Document Frequency (TF-IDF) values [16]. TF-IDF can be calculated based on the frequency of a term in a given page divided by the log of the frequency of the term in all pages.

In addition to employing machine learning techniques for phishing site identification, other methods exist. For example, Aburrous et al. [17] employed fuzzy logic to intelligently detect phishing websites. Wardman & Warner [18] proposed a novel idea that uses a hashing algorithm MD5 to identify phishing websites. They considered MD5 values of additional files found on a website (e.g., .js, .css, .jpg, .gif) for phish identification, with the assumption that many phishers would re-use the same files to create phishing websites. A similar assumption was made by Layton et al. [19], who proposed the use of Diffs to increase the ability of clustering algorithms to discover the differences between a phishing web page and the actual web page that is being impersonated. Fu et al. [20] used the earth mover's distance to measure the similarity of web page visual in identifying phishing web pages. They first converted images from a web page into low resolution images and used them as image signatures. After that, signature similarities between benign and phishing webpages were evaluated and malicious ones identified. Zhang et al. [11] presented a similar approach but considered both visual and textual contents, and applied a Bayesian classifier to detect phishing web pages. Liu et al. [21] used clustering of associated web pages to detect whether a given web page is a phishing page or not. A major issue with these methods is that they require processing of the original web page as a base comparison. This not only affects the system performance when detecting phishing web pages, but more importantly they require legitimate web pages as inputs given by the users. This brings doubts to the practicality of these methods, as some web domains may be unknown to the users.

Apart from phishing sites, various web domains and pages also perform malicious acts by distributing malware. Malware propagations are mostly in a stealth mode, where malicious programs would be downloaded to users' machines without their consent. This type of attacks is normally named as drive-by-download attack. An effective mitigation strategy is to detect an attack prior to loading a web domain or web page. This is typically done by reviewing the web code either manually or automatically, with the aim to identify malicious signatures. However, hackers are smart in obfuscating code to evade the code review process. Various studies that have employed machine learning techniques to detect drive-by download attacks exist. Cova et al. [22] introduced Wepawet, an online service that performs dynamic analysis on the web code using machine learning techniques. The dynamic analysis involves running the code in a sandboxed environment to record its behaviour. A machine learning algorithm is then trained to classify malicious sites. Canali et al. extended Wepawet with Prophiler [10] that performs quick static analysis using extracted features from the content of a web page with a decision tree algorithm. Rieck et al. [23] proposed Cujo, an automatic system that uses Support Vector Machines (SVMs) to detect drive-by-download attacks. Cujo performs both static and dynamic analyses after lexical tokens are extracted from the web code. Cujo uses similar features used by Wepawet and Prophiler. Curtsinger et al. [13] proposed Zozzle, an in-browser malware detection method. Zozzle uses Bayesian classification of hierarchical features of Javascript to identify syntax elements of malware. In addition to drive-by-download malware detection, several methods have been proposed to detect malware in a system. For example, Dinaburg et al. [24] proposed Ether, an analyser that uses hardware virtualisation extensions to analyse malware. Dinaburg et al. claimed that Ether remains transparent and does not induce any side effects that are unconditionally detectable by malware, as it resides outside of the target operating system environment. Other methods proposed to identify and analyse malware affecting system components include Justin [25] and PolyUnPack [26].

As can be seen from the review of related work presented, phishing and malware attacks are great threats to web users' privacy and security. While many studies in the literature have developed different kinds of approaches for identifying phishing and malware domains separately, we consider dealing with both attacks at the same time as crucial by building classification models with freely available phishing and malware related data features. To the best of our knowledge, our work is among the first in attempting to classify phishing and malware domains simultaneously.

## III. Models

In this section, we present the machine learning models used for comparison purposes. We first describe the single and ensemble models, and then explain how BPSO-based feature selection is incorporated into the single classifiers.

### A. Single Models

- An *Artificial Neural Network* (ANN) is a computational system composed of interconnected neurons, which exchange messages between each other [27]. A typical architecture of ANN consists of three layers: an input layer, which corresponds to the input features; an output layer, which denotes the response or output of the model; and a hidden layer that sits in between the input and output layers. To generate the final classifier that minimises the empirical errors, the network is trained by tuning the weight between each pair of connections.
- An *SVM* implements the structural risk minimisation principle by minimising an upper bound to the generalisation error [28]. The concept is to construct an optimal separating hyperplane in a high-dimensional feature space by maximising the margin between the hyperplane and the closest points of any class. This leads to excellent generalisation performance of SVM, which has been widely and successfully applied in many areas (e.g., see [29-31]).
- *C4.5*, introduced by Quinlan [32], is an algorithm that creates classifiers by building decision trees. It builds a tree from a dataset using the divide-and-conquer strategy. In case all samples of a set belong to the same class or the set is small, each leaf of the tree is labelled with the most frequent label. Otherwise, C4.5 chooses a feature with the highest gain ratio to split the dataset.
- *K-Nearest Neighbours* (KNN) classifies an unlabelled sample based on the k-nearest samples by assuming that those samples with similar properties will exist in close proximity [33]. The most frequent class label among the k-nearest neighbours is the predicted class label for the unlabelled sample.
- *Naïve Bayes* is a linear probabilistic classifier based on the Bayes theorem [34] with a simple assumption that each pair of features is independent given the context of the class. Although this assumption is clearly 'naïve' in most real-world applications, Naïve Bayes has worked quite well in many tasks such as document classification and spam filtering [35, 36].

### B. Ensemble Models

An ensemble approach typically combines a number of weak learners so that a strong learner with higher accuracy can be produced. Many studies, both theoretical and empirical, have shown that an ensemble of individual classifiers could outperform a single classifier (e.g., see [37, 38]). In this study, four well-known ensemble models are considered. They are briefly described as follows.

- *Bagging* (also known as bootstrap aggregation) constructs a set of weak learners by bootstrapping the samples [39]. Specifically, by taking many samples from the original dataset, bagging builds a separate classifier on each bootstrapped sample and generates the final classifier based on a majority vote strategy. Bagging can dramatically reduce the variance of weak learners like decision trees and lead to improved performance.
- *Adaboost*, which stands for Adaptive Boost, is one of the most prominent members of the Boost method [37]. Similar to bagging, Adaboost constructs a set of weak learners by manipulating the samples. Unlike bagging, which builds each learner on the bootstrapped sample independently, Adaboost is adaptive in the sense that each weak learner is built sequentially. By maintaining a set of adaptive weights, the subsequent learner pays more attention to those instances misclassified by previous classifiers. A composited classifier is generated by aggregating the weak learners through a weighted voting mechanism on the basis of their accuracy.

- *Random Forest* (RF), introduced by Breiman [40], is an improved version of bagging. It consists of many decision trees built on bootstrapped samples. The difference is that each decision tree is built with randomly selected subsets of features rather than the entire feature set as in bagging. The strategy of random selection of features enforces diversity without compromising the accuracy of individual learners, thus RF is able to improve the performance of bagging.

- The *Gradient Boosting Machine* (GBM) [41] improves boosting [37] as a method for function estimation from the perspective of optimisation. In a GBM, the learning procedure sequentially builds new base-learners with maximum gradient descent of the loss function.

### C. BPSO-based Feature Selection

Among the collected features (see Section IV.A) in this study, some may be less relevant or highly redundant compared to others. The presence of those features not only increases complexity but can decrease classification performance. Using feature selection to eliminate irrelevant and redundant features without decreasing the performance of a model is thus desirable (e.g., see [42-44]). BPSO [45], an efficient swarm intelligence technique, has been successfully used for feature selection in many classification problems (e.g., see [46, 47]). With this in mind, we apply it to the five single models described in Section III.A in order to improve their performance in identifying malicious web domains.

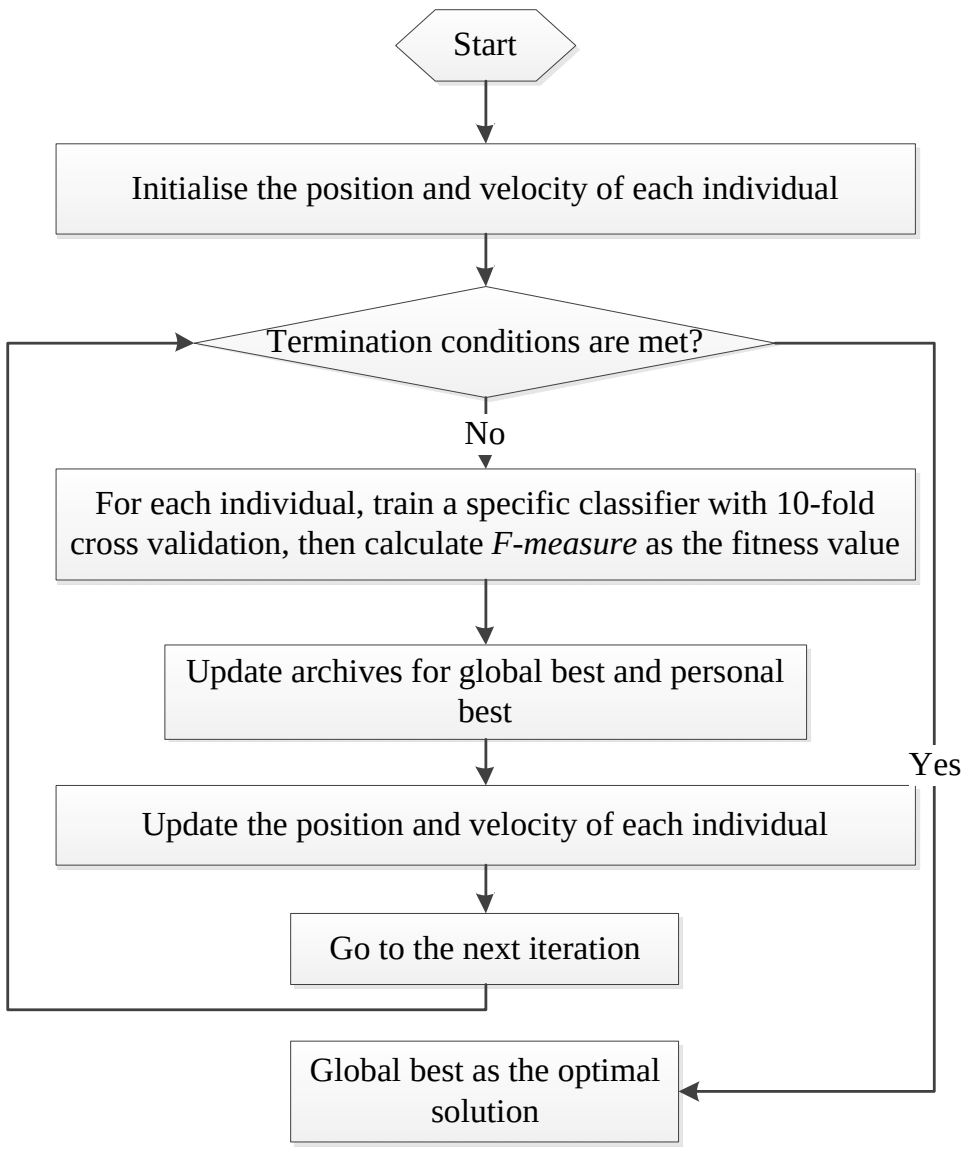

Fig. 1. A flowchart of BPSO-based feature selection

Fig. 1 shows the process of BPSO-based feature selection. An individual here is represented by a binary string corresponds to a feature subset. The length of each binary string equals to the number of available features, and a bit with the value of 1 (or 0) in the binary string denotes that the corresponding feature is selected (or excluded). To evaluate the fitness of an individual, a specific classifier is trained and the classification performance is calculated. In this study, for different classifiers with BPSO-based feature selection, the corresponding classifier is applied to calculate the fitness of an individual. By simulating social behaviour of birds flocking to a near-optimal position through interactions with others, the BPSO-based feature selection method can find a subset of features with improved performance for a specific classifier.

## IV. Experiments and Results

### A. Data Preparation

The objective of this paper is to improve the performance of machine learning in accurately identifying malicious web domains based on their popularity and performance. For this purpose, we collected malicious and benign live Internet web domains. Specifically, two datasets, one with malware only (Malware) and the other with both malware and phishing (MalwarePhishing) domains were collected to build and evaluate the classifiers. Each dataset consists of 2000 live Internet domains valid as at 31 July 2015. To eliminate the data imbalance issue, both datasets contain an equal amount of malicious and benign web domains. Out of the 1000 benign web domains, 250 records (25%) are popular web domains listed in the Alexa's top 500 Global Sites [48]. Malware web domains in the Malware dataset were obtained from MalwareDomain's blocklist [49] while phishing web domains in the combined MalwarePhishing dataset were obtained from PhishTank's blocklist [50]. Among the malware web domains are malware propagation domains, attackers' exploit domains, ransomware domains and high risk domains. Legitimate domains that are infected by malware or known to distribute them have also been included in the dataset. A data label, named "WebDomainLabel", is used to determine whether a web domain record is malicious or not.

In total, 16 data features were collected and used in the datasets. These data features are all in numerical values. A summary of the 16 data features is provided below:

1) Moz's Domain Authority denotes Moz's prediction [51] of how a website will perform in search engine rankings.
2) MozRank is a logarithmically scaled 10-point measure of global link popularity by Moz.
3) Moz's total backlinks represent the number of external backlinks received by a web domain. The more backlinks a web domain receives, the more popular the web domain is on the Internet.
4) Majestic's Citation Flow is a citation ranking given by Majestic SEO [52] that shows how influential a web domain is on the Internet based on citations by other domains.

5) Majestic's Trust Flow is a trust ranking given by Majestic SEO that shows how close those citations are to the trustworthy sources.
6) Majestic's backlinks represent the total number of external backlinks received by a web domain.
7) Majestic's reference domains represent the total number of external web domains that refer to a particular web domain.
8) Three data features from social authority metrics including
   a) total shares in Facebook,
   b) the total number of Twitter tweets, and
   c) the total number of +1 in Google plus.

   Social authority metrics enable us to find out the popularity of a web domain in various social network sites. The higher the social authority metric, the more popular the web domain is.
9) Google's Page Rank is a metric developed by Google that is used to rank web domains in Google's search engine results.
10) Google's Page Speed is a speed metric that shows the performance of a web domain to load.
11) Alexa's rank is a ranking given by Alexa that shows the popularity of a web domain on the Internet solely based on the number of visitors accessing the web domain. The higher the number of visitors, the more popular the web domain is.
12) Alexa's 1-month reach is the metric showing daily unique visitors in a one-month period.
13) Alexa's 3-month reach is the metric showing daily unique visitors in a three-month period.
14) Alexa's median load is the median load time of a web domain based on Alexa's survey with its proprietary algorithms.

Note that data obtained from these Internet sources may change or be updated over time, as web domains may gain or lose popularity and performance. It is therefore imperative to show the time-stamp of the collected data in our datasets. To avoid features in greater numeric ranges dominating those in smaller numeric ranges, all features have been scaled to the range of [0,1].

### B. *Evaluation Metrics*

For binary classification tasks such as to determine if a web domain is benign or malicious (malware or phishing ones), the performance of a classifier can be measured by using a confusion matrix (see Table I). As the cost of misclassifying a malicious web domain is much higher than that of a benign domain, "malicious" is labelled as the positive class. As can be seen in Table I, TP is the number of correctly classified malicious domains, TN is the number of correctly classified benign domains, FP is the number of predicted malicious domains that are actually benign ones, and FN is the number of wrongly classified benign domains that are actually malicious ones. Based on these four metrics, the performance of classification can be evaluated by the following measures: *Accuracy*, *Precision*, *Recall*, and *F-measure*. They are defined as follows:

$$Accuracy = \frac{TP + TN}{TP + TN + FP + FN} \quad (1)$$

$$Precision = \frac{TP}{TP + FP} \quad (2)$$

$$Recall = \frac{TP}{TP + FN} \quad (3)$$

$$F_measure = \frac{2 * Precision * Recall}{Precision + Recall} \quad (4)$$

Here, precision is the percentage of malicious domains over the classified malicious ones. Recall evaluates how well an algorithm has correctly classified the malicious domains. To compare the effectiveness by synthesising precision and recall, F-measure calculates the harmonic average value of precision and recall. Accuracy is a measure of how well the model has correctly classified the domains.

TABLE I. THE CONFUSION MATRIX FOR WEBSITE CLASSIFICATION

| **Actual** | **Predicted** | |
|---|---|---|
| | *Malicious* | *Benign* |
| *Malicious (Positive)* | TP (True Positive) | FN (False Negative) |
| *Benign (Negative)* | FP (False Positive) | TN (True Negative) |

### C. *Experimental Setups*

All experiments reported in this paper were executed in the R programming environment. The R packages 'e1071', 'nnet', 'class', 'RWeka', 'klaR', 'randomForest', 'gbm' and 'adabag' were applied to implement the machine learning algorithms. All of these packages are available from the Comprehensive R Archive Network repository (http://cran.r-project.org/). The BPSO-based feature selection technique was implemented in R by ourselves. Some adjustable parameters in each of the classifiers were tuned by maximising *F-measure* using 5-fold cross validation. Specifically, the tuned parameters include: SVM (RBF Kernel, gamma = 0.5, Cost = 8), ANN (size = 5, maxit = 2000), C4.5 (C = 0.05), KNN (K = 10), RF (ntree = 1000), GBM (n.trees = 5000, distribution = 'bernoulli', n.minobsinnode = 10), Adaboost (mfinal = 100, maxdepth = 3), bagging (mfinal = 100, maxdepth = 5). For the BPSO-based feature selection technique, its swarm size was 30 and maximum iteration was 500. Both cognitive and interaction coefficients were set to 2.0.

### D. *Experimental Results and Dicussion*

To identify the malware and phishing websites based on popularity and performance data, nine well-studied machine learning algorithms, which include five single models (i.e., SVM, ANN, KNN, C4.5 and Naïve Bayes) and four ensemble models (i.e., RF, GBM, Adaboost and bagging), were investigated. Cross validation was first applied for parameter tuning so that the optimal value of each parameter can be determined. Then, 10-fold cross validation with random permutation was done to estimate the performance of each classifier. Furthermore, the average performance and standard deviation were estimated by repeating the cross validation 10 times so as to reduce bias in the experiments.

Table II and Table III report the performance of the nine classifiers on the Malware and combined MalwarePhishing datasets, respectively. As can be seen from Table II, among the five single models, four of them consistently perform well with the values of different metrics larger than 0.88. On the contrary, Bayes performs very badly in three of the metrics. Although Bayes gains the highest recall with the value of 0.9849, which means it can correctly identify malware and phishing websites with the highest percentage, it misclassifies a large number of benign domains as malicious ones in terms of accuracy and precision. From the viewpoint of F-measure, which is considerably the most important measure in a problem of this nature, the SVM has the highest value of 0.9123, followed by KNN (0.9113), ANN (0.9089) and C4.5 (0.9044). Moving to the ensemble models, RF performs the best with F-measure of 0.9393. Compared to the single models, the ensembles are consistently better across all four metrics, with an exception of bagging, which is slightly worse than the five single models in terms of recall. Similar results can be observed in Table III.

From the above, it is clear that the ensemble approach that combines a number of weak learners could generate better performance. The four ensemble models considered in this study have all been formed by combining several weak learners. The superiority of ensemble models over single ones observed in this study is consistent with findings from past studies [37, 38]. Among the four ensemble models, RF, the GBM and Adaboost have a built-in feature selection process, which improves the performance further. Feature selection can identify a powerful predictive subset of variables by eliminating noisy, irrelevant and redundant inputs without degrading the performance of a model. It is thus often applied for single models. For a fairer comparison, we incorporated feature selection into the single models to see if their performance could be improved.

Table IV and Table V show the results of feature selection based single models on the Malware and combined MalwarePhishing datasets, respectively. In comparison with the original single models, these feature selection based models have better performances. Especially for the Bayes approach, substantial improvement in performance (except F-measure) can be observed. Nevertheless, the feature selection based single models are still performing slightly worse than the ensembles.

Fig. 2 and Fig. 3 illustrate the average rank of mean performance and standard deviation of performance of different classifiers on the two datasets. The average ranks were computed through the Friedman test. Based on the Friedman test, the model with the best performance is ranked as 1 while the worst model is ranked as 14 (there are 14 models in total). From Fig. 2, we can clearly conclude that the ensemble models have lowest average ranks, followed by the feature selection based single models and finally the single models.

TABLE II. COMPARISON OF PERFORMANCE OF SEVERAL CLASSIFIERS: THE MALWARE DATASET

| | | **Accuracy** | **Precision** | **Recall** | **F-measure** |
|---|---|---|---|---|---|
| Single | SVM | 0.9109 | 0.8945 | 0.9314 | 0.9123 |
| | ANN | 0.9076 | 0.8930 | 0.9261 | 0.9089 |
| | KNN | 0.9097 | 0.8925 | 0.9316 | 0.9113 |
| | C4.5 | 0.9023 | 0.8813 | 0.9299 | 0.9044 |
| | Bayes | 0.7280 | 0.6505 | 0.9849 | 0.7830 |
| Ensemble | RF | 0.9387 | 0.9293 | 0.9499 | 0.9393 |
| | GBM | 0.9356 | 0.9268 | 0.9457 | 0.9360 |
| | Adaboost | 0.9356 | 0.9294 | 0.9430 | 0.9359 |
| | bagging | 0.9150 | 0.9085 | 0.9224 | 0.9150 |

TABLE III. COMPARISON OF PERFORMANCE OF SEVERAL CLASSIFIERS: THE MALWAREPHISHING DATASET

| | | **Accuracy** | **Precision** | **Recall** | **F-measure** |
|---|---|---|---|---|---|
| Single | SVM | 0.9155 | 0.9008 | 0.9344 | 0.9168 |
| | ANN | 0.9095 | 0.8985 | 0.9240 | 0.9104 |
| | KNN | 0.9145 | 0.8998 | 0.9332 | 0.9157 |
| | C4.5 | 0.9086 | 0.8943 | 0.9278 | 0.9100 |
| | Bayes | 0.7478 | 0.6680 | 0.9856 | 0.7956 |
| Ensemble | RF | 0.9425 | 0.9342 | 0.9525 | 0.9430 |
| | GBM | 0.9433 | 0.9340 | 0.9544 | 0.9438 |
| | Adaboost | 0.9396 | 0.9350 | 0.9456 | 0.9399 |
| | bagging | 0.9195 | 0.9130 | 0.9281 | 0.9200 |

TABLE IV. COMPARISON OF PERFORMANCE OF FEATURE SELECTION BASED CLASSIFIERS: THE MALWARE DATASET

| | | **Accuracy** | **Precision** | **Recall** | **F-measure** |
|---|---|---|---|---|---|
| Feature selection | SVM | 0.9211 | 0.9005 | 0.9470 | 0.9229 |
| | ANN | 0.9153 | 0.9018 | 0.9325 | 0.9166 |
| | KNN | 0.9191 | 0.8976 | 0.9468 | 0.9212 |
| | C4.5 | 0.9148 | 0.8935 | 0.9417 | 0.9165 |
| | Bayes | 0.8865 | 0.8605 | 0.9227 | 0.8901 |

TABLE V. COMPARISON OF PERFORMANCE OF FEATURE SELECTION BASED CLASSIFIERS: THE MALWAREPHISHING DATASET

| | | **Accuracy** | **Precision** | **Recall** | **F-measure** |
|---|---|---|---|---|---|
| Feature selection | SVM | 0.9210 | 0.9150 | 0.9440 | 0.9300 |
| | ANN | 0.9135 | 0.8988 | 0.9331 | 0.9150 |
| | KNN | 0.9177 | 0.9004 | 0.9399 | 0.9193 |
| | C4.5 | 0.9185 | 0.9103 | 0.9287 | 0.9187 |
| | Bayes | 0.8955 | 0.8738 | 0.9252 | 0.8982 |

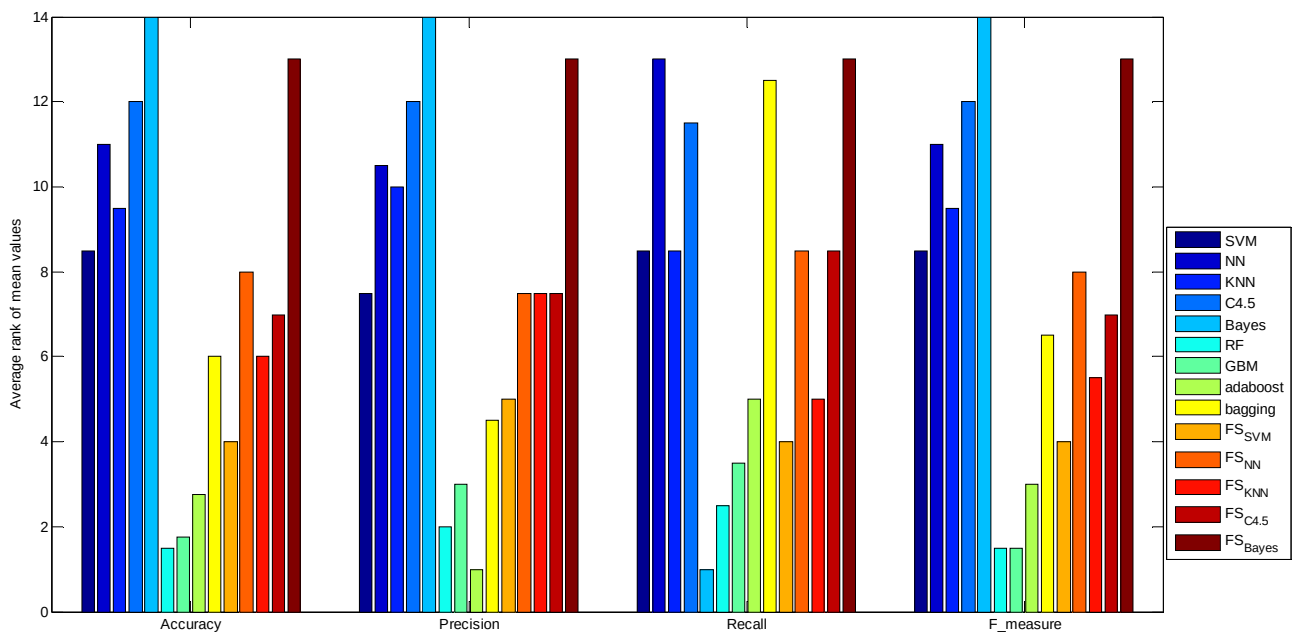

Fig. 2. The average rank of mean performance of different classifiers over the two datasets

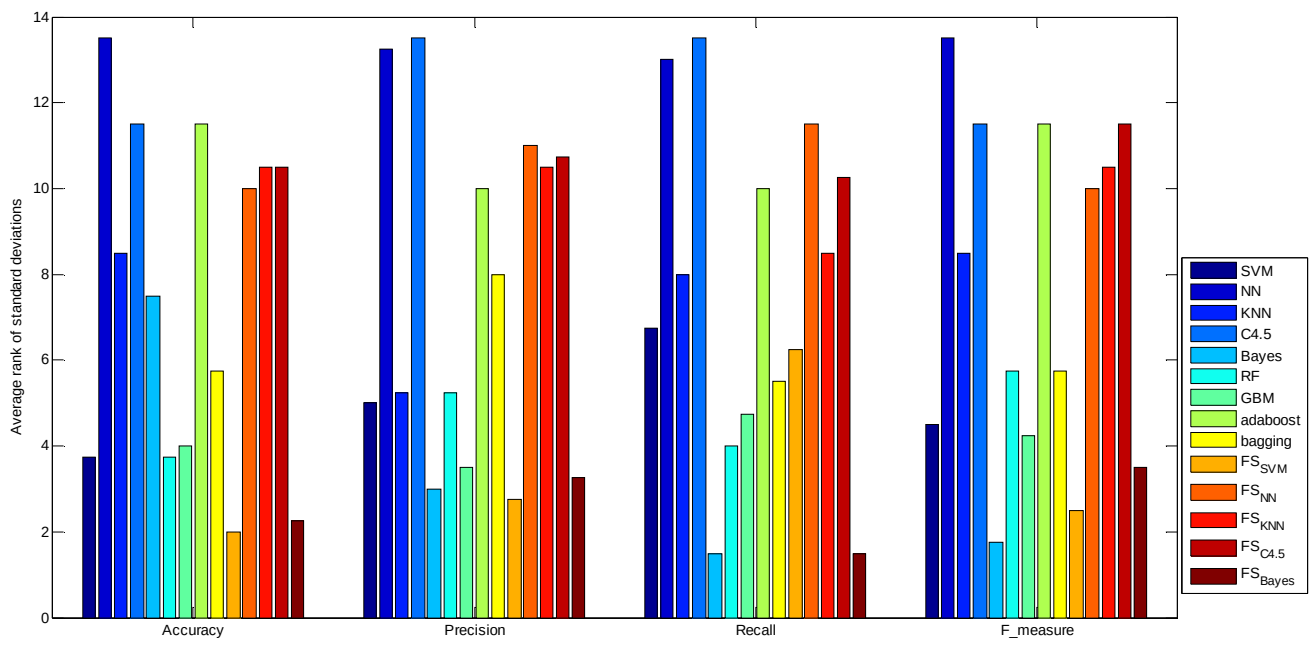

Fig. 3. The average rank of standard deviation of performance of different classifiers over the two datasets

TABLE VI. WILCOXON'S SIGNED-RANK TEST BETWEEN EACH PAIR OF CLASSIFIERS ON THE MALWARE DATASET (F-MEASURE)

| | ANN | KNN | C4.5 | Bayes | RF | GBM | Adaboost | bagging | FS_SVM | FS_ANN | FS_KNN | FS_C4.5 | FS_Bayes |
|---|---|---|---|---|---|---|---|---|---|---|---|---|---|
| SVM | 3.71E-02 | **4.92E-01** | 1.95E-03 | 1.95E-03 | 1.95E-03 | 1.95E-03 | 1.95E-03 | 3.91E-03 | 1.95E-03 | **1.93E-01** | 1.37E-02 | **2.32E-01** | 1.95E-03 |
| ANN | | **1.93E-01** | 1.95E-03 | 1.95E-03 | 1.95E-03 | 1.95E-03 | 1.95E-03 | 3.91E-03 | 1.95E-03 | **4.92E-01** | 1.95E-03 | **6.25E-01** | 1.95E-03 |
| KNN | | | 1.95E-03 | 1.95E-03 | 1.95E-03 | 1.95E-03 | 1.95E-03 | 1.37E-02 | 1.95E-03 | **6.25E-01** | 1.37E-02 | **4.92E-01** | 1.95E-03 |
| C4.5 | | | | 1.95E-03 | 1.95E-03 | 1.95E-03 | 1.95E-03 | 1.95E-03 | 1.95E-03 | 3.91E-03 | 1.95E-03 | 2.73E-02 | 1.95E-03 |
| Bayes | | | | | 1.95E-03 | 1.95E-03 | 1.95E-03 | 1.95E-03 | 1.95E-03 | 1.95E-03 | 1.95E-03 | 1.95E-03 | 1.95E-03 |
| RF | | | | | | 1.95E-03 | 1.95E-02 | 1.95E-03 | 1.95E-03 | 1.95E-03 | 1.95E-03 | 1.95E-03 | 1.95E-03 |
| GBM | | | | | | | **1.00E+00** | 1.95E-03 | 1.95E-03 | 5.89E-03 | 1.95E-03 | 1.95E-03 | 1.95E-03 |
| Adaboost | | | | | | | | 1.95E-03 | 1.95E-03 | 1.95E-03 | 1.95E-03 | 1.95E-03 | 1.95E-03 |
| bagging | | | | | | | | | **1.05E-01** | 3.91E-03 | **8.46E-01** | 1.95E-02 | 1.95E-03 |
| FS_SVM | | | | | | | | | | 1.95E-03 | 3.71E-02 | 9.77E-03 | 1.95E-03 |
| FS_ANN | | | | | | | | | | | 9.77E-03 | **1.00E+00** | 1.95E-03 |
| FS_KNN | | | | | | | | | | | | 8.40E-02 | 1.95E-03 |
| FS_C4.5 | | | | | | | | | | | | | 1.95E-03 |

TABLE VII. WILCOXON'S SIGNED-RANK TEST BETWEEN EACH PAIR OF CLASSIFIERS ON THE MALWAREPHISHING DATASET (F-MEASURE)

| | ANN | KNN | C4.5 | Bayes | RF | GBM | Adaboost | bagging | FS_SVM | FS_ANN | FS_KNN | FS_C4.5 | FS_Bayes |
|---|---|---|---|---|---|---|---|---|---|---|---|---|---|
| SVM | 5.86E-03 | **1.93E-01** | 1.95E-03 | 1.95E-03 | 1.95E-03 | 1.95E-03 | 1.95E-03 | 1.95E-03 | 1.95E-03 | 1.95E-03 | **4.32E-01** | **3.22E-01** | 1.95E-03 |
| ANN | | 1.95E-03 | **8.46E-01** | 1.95E-03 | 1.95E-03 | 1.95E-03 | 1.95E-03 | 1.95E-03 | 1.95E-03 | **2.75E-01** | 1.95E-03 | 3.91E-03 | 1.95E-03 |
| KNN | | | 5.86E-03 | 1.95E-03 | 1.95E-03 | 1.95E-03 | 1.95E-03 | 1.95E-03 | 1.95E-03 | 1.37E-02 | **2.32E-01** | **1.60E-01** | 1.95E-03 |
| C4.5 | | | | 1.95E-03 | 1.95E-03 | 1.95E-03 | 1.95E-03 | 1.95E-03 | 1.95E-03 | **1.60E-01** | 3.91E-03 | 3.91E-03 | 1.95E-03 |
| Bayes | | | | | 1.95E-03 | 1.95E-03 | 1.95E-03 | 1.95E-03 | 1.95E-03 | 1.95E-03 | 1.95E-03 | 1.95E-03 | 1.95E-03 |
| RF | | | | | | 2.60E-01 | 2.73E-02 | 1.95E-03 | 1.95E-03 | 1.95E-03 | 1.95E-03 | 1.95E-03 | 1.95E-03 |
| GBM | | | | | | | 9.77E-03 | 1.95E-03 | 1.95E-03 | 1.95E-03 | 1.95E-03 | 1.95E-03 | 1.95E-03 |
| Adaboost | | | | | | | | 1.95E-03 | 1.95E-03 | 1.95E-03 | 1.95E-03 | 1.95E-03 | 1.95E-03 |
| bagging | | | | | | | | | 1.95E-03 | 1.95E-03 | **8.40E-02** | **2.32E-01** | 1.95E-03 |
| FS_SVM | | | | | | | | | | 1.95E-03 | 1.95E-03 | 1.95E-03 | 1.95E-03 |
| FS_ANN | | | | | | | | | | | 9.77E-03 | 1.95E-02 | 1.95E-03 |
| FS_KNN | | | | | | | | | | | | **6.95E-01** | 1.95E-03 |
| FS_C4.5 | | | | | | | | | | | | | 1.95E-03 |

To further verify the significance between performances of different classifiers, a non-parametric test - the Wilcoxon signed-rank test - was applied to examine the statistical differences between the results obtained by each of the classifiers. Table VI and Table VII show the p-values of these statistical tests based on F-measure. In these tables, the p-values that are greater than the significant level (i.e., > 0.05) have been highlighted in bold. From the tables, we can conclude that: (1) in terms of F-measure on the two datasets, most of the pairwise classifiers are significantly different with only a small number of exceptions; (2) the ensemble models statistically outperform both single models and the feature selection based single models except that there are no statistical differences between bagging and FS_SVM as well as bagging and FS_KNN on the Malware dataset, and bagging and FS_KNN as well as bagging and FS_C4.5 on the MalwarePhishing dataset; and (3) by comparing the feature selection based single models and single models, except for ANN (on both datasets) and KNN (on the MalwarePhishing dataset), the feature selection process statistically improves the results of single models.

## V. Conclusions and Future Work

In this paper, we have presented a study to investigate the performance of machine learning techniques for identifying malware and phishing web domains using online popularity and performance data. Sixteen data features that represent the popularity and performance of web domains were crawled, and nine well-known machine learning techniques, including both single and ensemble models, were examined. Experimental results show that the ensemble models significantly outperform those single models investigated. In addition, a BPSO based feature selection method was incorporated to the single models and the results show that it can effectively improve the performance of single models for malicious domain classification. This further indicates the feasibility of employing the widely available Internet data about web domains to accurately differentiate malicious web domains from benign ones. Unlike the existing methods that extract features from the content of malicious web domains and compare them with the benign ones', our approach does not require this tedious process. Moreover, our proposed approach can identify the malicious web domains regardless of whether they perform phishing attacks or distribute malware, which is essential in real world settings where a malicious web domain can be a phishing site or would distribute malware, or both.

This study motivates future work in exploring the inclusion of more online data to improve the predictive performance of machine learning techniques. Besides that, imbalanced class distribution, a challenging problem in the area of malicious domain identification, should be addressed. Finally, an automated filter that can extract features and build identification models automatically will be investigated.

## Acknowledgment

We would like to acknowledge support from the Natural Science Foundation of China under Project No. 71571080, the Fundamental Research Funds for the Central Universities (2014QN205-HUST), the China Postdoctoral Science Foundation (2015M582280), a grant from the Modern Information Management Research Center at Huazhong University of Science and Technology, and the auDA Foundation. We would also like to acknowledge the in-kind support from Majestic SEO (https://majestic.com/), which allowed us to obtain the Majestic data free of charge.

## References

[1] Netcraft Extension. (1 December 2015). Available: http://toolbar.netcraft.com/

[2] SpoofGuard. (1 December 2015). Available: https://crypto.stanford.edu/SpoofGuard/

[3] Y. Zhang, S. Egelman, L. Cranor, and J. Hong, "Phinding phish: Evaluating anti-phishing tools," in *Proceedings of the 14th Annual Network & Distributed System Security Symposium (NDSS 2007),* San Diego, CA, 2006, pp. 1-17.

[4] A. Blum, B. Wardman, T. Solorio, and G. Warner, "Lexical feature based phishing URL detection using online learning," in *Proceedings of the 3rd ACM Workshop on Artificial Intelligence and Security*, 2010, pp. 54-60.

[5] J. Ma, L. K. Saul, S. Savage, and G. M. Voelker, "Identifying suspicious URLs: An application of large-scale online learning," in *Proceedings of the 26th Annual International Conference on Machine Learning*, 2009, pp. 681-688.

[6] J. Ma, L. K. Saul, S. Savage, and G. M. Voelker, "Beyond blacklists: Learning to detect malicious web sites from suspicious URLs," in *Proceedings of the 15th ACM SIGKDD International Conference on Knowledge Discovery and Data Mining*, 2009, pp. 1245-1254.

[7] S. Garera, N. Provos, M. Chew, and A. D. Rubin, "A framework for detection and measurement of phishing attacks," in *Proceedings of the 2007 ACM Workshop on Recurring Malcode*, 2007, pp. 1-8.

[8] G. Xiang, J. Hong, Carolyn P Rose, and L. Cranor, "Cantina+: A feature-rich machine learning framework for detecting phishing web sites," *ACM Transactions on Information and System Security (TISSEC),* vol. 14, 2011.

[9] Y. Zhang, J. Hong, and L. Cranor, "CANTINA: A content-based approach to detecting phishing web sites," in *Proceedings of the 16th International Conference on World Wide Web (WWW'07)*, 2007, pp. 639-648.

[10] D. Canali, M. Cova, G. Vigna, and C. Kruegel, "Prophiler: A fast filter for the large-scale detection of malicious web pages," in *Proceedings of the International World Wide Web Conference (WWW'11),* Hyderabad, India, 2011, pp. 197-206.

[11] H. Zhang, G. Liu, T. Chow, and W. Liu, "Textual and visual contentbased anti-phishing: A bayesian approach," *IEEE Transactions on Neural Networks,* vol. 22, pp. 1532-1546, 2011.

[12] C. Ludl, S. McAllister, E. Kirda, and C. Kruegel, "On the effectiveness of techniques to detect phishing sites," in *Proceedings of the 4th International Conference on Detection of Intrusions and Malware, and Vulnerability Assessment (DIMVA'07),* Lucerne, Switzerland, 2007, pp. 20-39.

[13] C. Curtsinger, B. Livshits, B. Zorn, and C. Seifert, "Zozzle: Fast and precise in-browser javascript malware detection," in *Proceedings of the 20th USENIX conference on Security (SEC 11),* San Fransisco, CA, 2011, pp. 33-48.

[14] N. Sanglerdsinlapachai and A. Rungsawang, "Using domain top-page similarity feature in machine learning-based web phishing detection," in *Proceedings of the 3rd International*

*Conference on Knowledge Discovery and Data Mining*, Phuket, Thailand, 2010, pp. 187-190.

[15] C. Whittaker, B. Ryner, and M. Nazif, "Large-scale automatic classification of phishing pages," in *Proceedings of the 17th Annual Network and Distributed System Security Symposium (NDSS 2010)*, 2010, pp. 1-14.

[16] G. Salton and M. J. McGill, *Introduction to Modern Information Retrieval*. New York, NY, USA: McGraw-Hill, Inc., 1983.

[17] M. Aburrous, M. A. Hossain, F. Thabatah, and K. Dahal, "Intelligent phishing website detection system using fuzzy techniques," in *Proceedings of the 3rd IEEE International Conference on Information and Communication Technologies: From Theory to Applications (ICTTA 2008)*, Damascus, 2008, pp. 1-6.

[18] B. Wardman and G. Warner, "Automating phishing website identification through deep md5 matching," in *eCrime Researchers Summit*, Atlanta, GA, 2008, pp. 1-8.

[19] R. Layton, S. Brown, and P. Watters, "Using differencing to increase distinctiveness for phishing website clustering," in *Symposia and Workshops on Ubiquitous, Autonomic and Trusted Computing*, Brisbane, QLD, 2009, pp. 488-492.

[20] A. Y. Fu, L. Wenyin, and X. Deng, "Detecting phishing web pages with visual similarity assessment based on earth mover's distance," *IEEE Transactions Dependable Secure Computation*, vol. 3, pp. 301-311, 2006.

[21] G. Liu, B. Qiu, and L. Wenyin, "Automatic detection of phishing target from phishing webpage," in *Proceedings of the 20th International Conference on Pattern Recognition (ICPR 2010)*, 2010, pp. 4153-4156.

[22] M. Cova, C. Kruegel, and G. Vigna, "Detection and analysis of drive-by-download attacks and malicious javascript code," in *Proceedings of the 19th International Conference on World Wide Web (WWW 2010)*, Raleigh, North Carolina, USA, 2010, pp. 281-290.

[23] K. Rieck, T. Krueger, and A. Dewald, "Cujo: Efficient detection and prevention of drive-by-download attacks," in *Proceedings of the 26th Annual Computer Security Applications Conference*, Orlando, Florida, 2010, pp. 31-39.

[24] A. Dinaburg, P. Royal, M. I. Sharif, and W. Lee, "Ether: Malware analysis via hardware virtualization extensions," in *Proceedings of the ACM Conference on Computer and Communications Security (CCS)*, 2008, pp. 51-62.

[25] F. Guo, P. Ferrie, and T.-C. Chiueh, "A study of the packer problem and its solutions," in *Proceedings of the 11th International Symposium on Recent Advances In Intrusion Detection (RAID)*, 2008, pp. 98-115.

[26] P. Royal, M. Halpin, D. Dagon, R. Edmonds, and W. Lee, "Polyunpack: Automating the hidden-code extraction of unpack-executing malware," in *Proceedings of the 22nd Annual Computer Security Applications Conference (ACSAC)*, 2006, pp. 289-300.

[27] R. Chiong, F. Neri, and R. I. McKay, "Nature that breeds solutions," in *Nature-inspired informatics for intelligent applications and knowledge discovery: Implications in business, science and engineering*, R. Chiong ed.: IGI Global, 2010, pp. 1-24.

[28] V. N. Vapnik, *The Nature of Statistical Learning Theory*. New York: Springer, 1995.

[29] S. L. Lo, D. Cornforth, and R. Chiong, "Identifying the high-value social audience from twitter through text-mining methods," in *Proceedings of the 18th Asia Pacific Symposium on Intelligent and Evolutionary Systems (IES 2014)*, Springer International Publishing, 2014, pp. 325-339.

[30] S. Tong and D. Koller, "Support vector machine active learning with applications to text classification," *The Journal of Machine Learning Research*, vol. 2, pp. 45-66, 2002.

[31] Z. Y. Hu, Y. K. Bao, and T. Xiong, "Comprehensive learning particle swarm optimization based memetic algorithm for model selection in short-term load forecasting using support vector regression," *Applied Soft Computing*, vol. 25, pp. 15-25, 2014.

[32] J. R. Quinlan, *C4. 5: Programs for Machine Learning*. Morgan Kaufmann, 1993.

[33] K. Fukunaga and P. M. Narendra, "A branch and bound algorithm for computing k-nearest neighbors," *IEEE Transactions on Computers*, vol. 100, pp. 750-753, 1975.

[34] M. Bayes and M. Price, "An essay towards solving a problem in the doctrine of chances. By the late Rev. Mr. Bayes, F. R. S. communicated by Mr. Price, in a letter to John Canton, A. M. F. R. S.," *Philosophical Transactions*, vol. 53, pp. 370-418, 1763.

[35] R. Chiong and B. T. Lau, "A hybrid naive bayes approach for information filtering," in *Proceedings of the 3rd IEEE Conference on Industrial Electronics and Applications (ICIEA 2008)*, Singapore, 2008, pp. 1003-1007.

[36] V. Metsis, I. Androutsopoulos, and G. Paliouras, "Spam filtering with naive bayes-which naive bayes?," in *Proceedings of the 3rd Conference on Email and Anti-Spam (CEAS 2006)*, Mountain View, California USA, 2006, pp. 27-28.

[37] Y. Freund and R. E. Schapire, "A decision-theoretic generalization of on-line learning and an application to boosting," *Journal of Computer and System Sciences*, vol. 55, pp. 119-139, 1997.

[38] S. L. Lo, R. Chiong, and D. Cornforth, "Using support vector machine ensembles for target audience classification on twitter," *PLoS ONE*, vol. 10, e0122855, 2015.

[39] L. Breiman, "Bagging predictors," *Machine Learning*, vol. 24, pp. 123-140, 1996.

[40] L. Breiman, "Random forests," *Machine Learning*, vol. 45, pp. 5-32, 2001.

[41] J. H. Friedman, "Greedy function approximation: A gradient boosting machine," *Annals of Statistics*, pp. 1189-1232, 2001.

[42] Z. Y. Hu, Y. K. Bao, T. Xiong, and R. Chiong, "Hybrid filter-wrapper feature selection for short-term load forecasting," *Engineering Applications of Artificial Intelligence*, vol. 40, pp. 17-27, 2015.

[43] Z. Y. Hu, Y. K. Bao, R. Chiong, and T. Xiong, "Mid-term interval load forecasting using multi-output support vector regression with a memetic algorithm for feature selection," *Energy*, vol. 84, pp. 419-431, 2015.

[44] M. Abedini, M. Kirley, and R. Chiong, "Incorporating feature ranking and evolutionary methods for the classification of high-dimensional DNA microarray gene expression data," *Australasian Medical Journal*, vol. 6, pp. 272-279, 2013.

[45] J. Kennedy and R. C. Eberhart, "A discrete binary version of the particle swarm algorithm," in *Proceeding of the International Conference on Systems, Man, and Cybernetics*, 1997, pp. 4104-4108.

[46] L.-Y. Chuang, C.-H. Yang, and J.-C. Li, "Chaotic maps based on binary particle swarm optimization for feature selection," *Applied Soft Computing*, vol. 11, pp. 239-248, 2011.

[47] L. Cervante, B. Xue, M. Zhang, and L. Shang, "Binary particle swarm optimisation for feature selection: A filter based approach," in *Proceedings of the IEEE Congress on Evolutionary Computation (CEC 2012)*, Brisbane, Australia, 2012, pp. 1-8.

[48] Alexa. (31 July 2015). Available: http://www.alexa.com/topsites

[49] MalwareDomains. (15 June 2015). Available: http://www.malwaredomains.com/

[50] Phishtank. (15 June 2015). Available: https://www.phishtank.com/

[51] Moz. (15 June 2015). Available: https://moz.com/researchtools/ose/

[52] Majestic. (10 June 2015). Available: https://majestic.com/